\documentclass[a4paper]{jpconf}
\usepackage{graphicx}
\begin{document}
\title{Correlations of Electrons from Heavy Flavor Decay with Hadrons in Au+Au and p+p Collisions}

\author{Anne M. Sickles, for the PHENIX Collaboration}

\address{Brookhaven National Laboratory, Upton, NY}

\ead{anne@bnl.gov}

\begin{abstract}
Measurements of electrons from the decay of open-heavy flavor mesons have 
shown that the yields are suppressed in Au+Au collisions compared to 
expectations from binary-scaled p+p collisions. These measurements 
indicate that charm and bottom quarks interact with the hot-dense matter 
produced in heavy-ion collisions much more than expected. Here we extend 
these studies to two-particle correlations where one particle is an 
electron from the decay of a heavy-flavor meson and the other is a 
charged hadron from either the decay of the heavy meson or from jet 
fragmentation. These measurements provide more detailed information 
about the interactions between heavy quarks and the matter, such as 
whether the modifcation of the away-side-jet shape seen in hadron-hadron 
correlations is present when the trigger particle is from heavy-meson 
decay and whether the overall level of away-side-jet suppression is 
consistent. We statistically subtract correlations of electrons arising 
from background sources from the inclusive electron-hadron correlations 
and obtain two-particle azimuthal correlations at $\sqrt{s_{NN}}$ =200
 GeV between 
electrons from heavy-flavor decay with charged hadrons in p+p and also 
first results in Au+Au collisions. We find the away-side-jet shape and 
yield to be modified in Au+Au collisions compared to p+p collisions.
\end{abstract}

\section{Introduction}
One of the main goals of heavy ion physics is to understand
the interactions between the produced matter (commonly understood
as a strongly couple quark gluon plasma, sQGP) and fast partons.
Measurements of the nuclear modification factor, $R_{AA}$, of 
$\pi^0$s~\cite{ppg080} show a large suppression which is 
understood as resulting from parton-matter interactions throughout
the time evolution of the matter.  Further studies, in which a trigger
$\pi^0$ is correlated with other hadrons in the event have shown a strong
suppression of back-to-back jet-like correlations~\cite{ppg106}.

These results are qualitatively as expected from energy-loss models.
However,
heavy charm and bottom quarks have the ability, via their large mass,
to provide additional constraints on the mechanisms by which the partons
interact with the matter.  Notably, heavy guarks are expected to suffer
less energy loss via gluon radiation in the matter because of the
dead cone effect~\cite{dead_cone}.  

Heavy quarks in heavy ion collisions have largely been studied via 
single electrons which come from the decays of $D$ and $B$ mesons which
carry the original heavy quark.  The $R_{AA}$ of these electrons has been shown
to be significantly below 1 and comparable to that of $\pi^0$s~\cite{ppg066}.
Measurements of single electrons are sensitive to both charm and bottom quarks,
with a mixture that depends on the electron $p_T$~\cite{ppg094,star_cbprl}.

In order to further study these effects we have performed two-particle correlations of
electrons from heavy flavor decay
with other  hadrons, in the same manner as has been done for $\pi^0$s.
This can provide  a more detailed view of the interactions between heavy quarks and
the sQGP.  However, there are additional complications when considering heavy 
quarks, thus baseline measurements in p+p collisions are vital.  The results shown
here have been published as Ref.~\cite{ppg112}.

\section{Method}
Experimentally, separating the correlations of heavy flavor electrons from
those of non-heavy flavor electrons (such as those from photon conversions and
Dalitz decays) is necessary.  The procedure used is outlined in Ref.~\cite{ppg112}.
In general, a precise knowledge of the relative yields of heavy flavor and non-heavy
flavor sources is needed.  With that and a measurement of the correlations of non-heavy
flavor electrons it is possible to statistically subtract the non-heavy
flavor correlations from the total electron correlations (which include both
the heavy flavor and non-heavy flavor correlations).  This is 
analogous to the procedure used to separate direct photon triggered correlations
from those triggered by decay photons in Ref.~\cite{ppg090}.



\section{Results}
\begin{figure*}[ht]
\begin{minipage}{0.48\linewidth}
\includegraphics[width=0.99\linewidth]{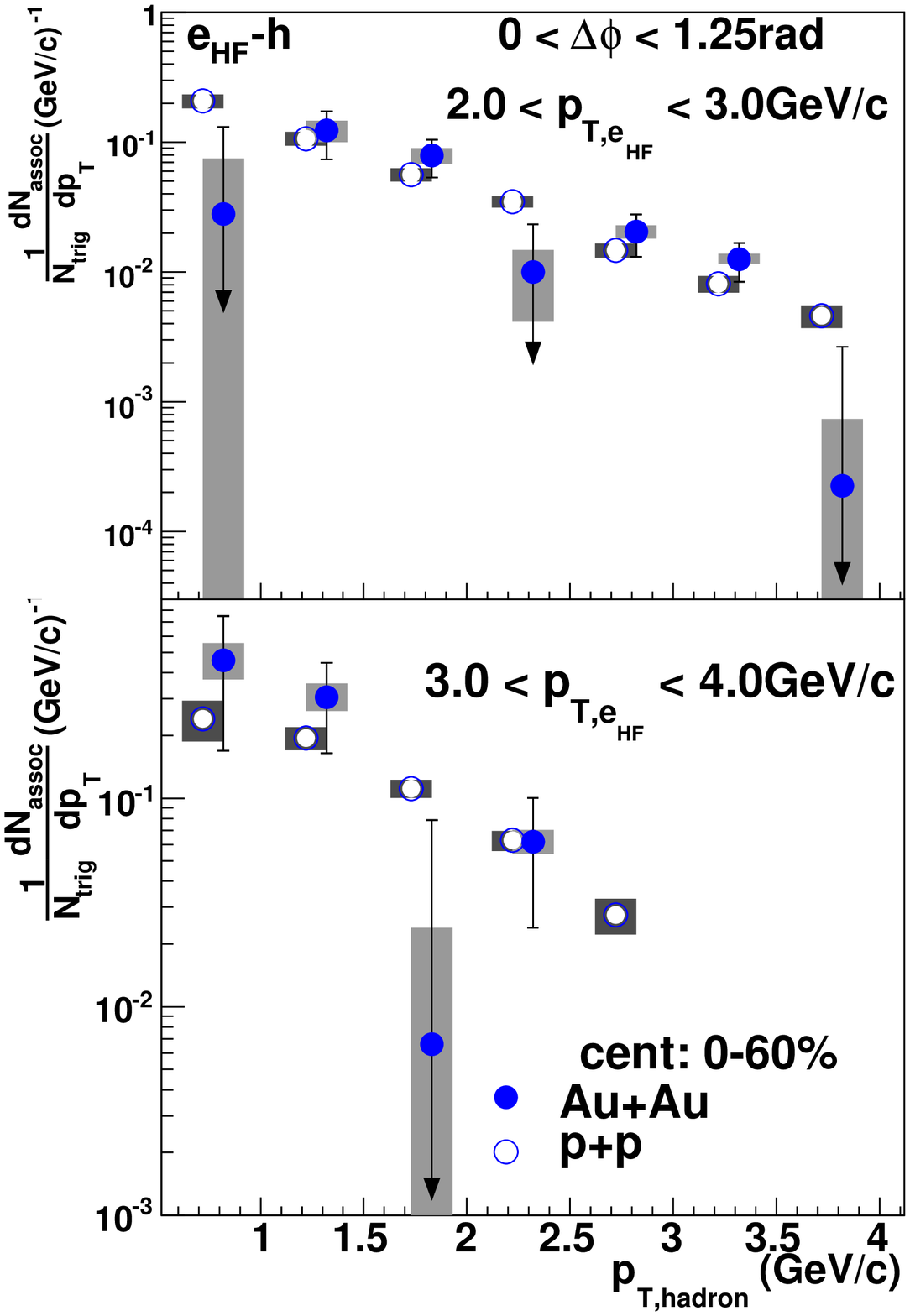}
\caption{\label{nearyield_auau} (color online) 
Near-side (0$<\Delta\phi<$1.25 rad) integrated yield for 
Au+Au (solid circles) and $p$+$p$ collisions (open circles) for 
2.0$<p_{T,e}<$3.0~GeV/$c$ (top panel) and 3.0$<p_{T,e}<$4.0~GeV/$c$ 
(bottom panel) as a function of the associated hadron $p_T$.  The 
overall normalization uncertainty of 9.4\% in Au+Au and 7.9\% in 
$p$+$p$ is not shown.  Points are slightly shifted horizontally for 
clarity.  Figure is from Ref.~\cite{ppg112}.}
\end{minipage}%
\hspace{0.5cm}
\begin{minipage}{0.48\linewidth}
\includegraphics[width=0.99\linewidth]{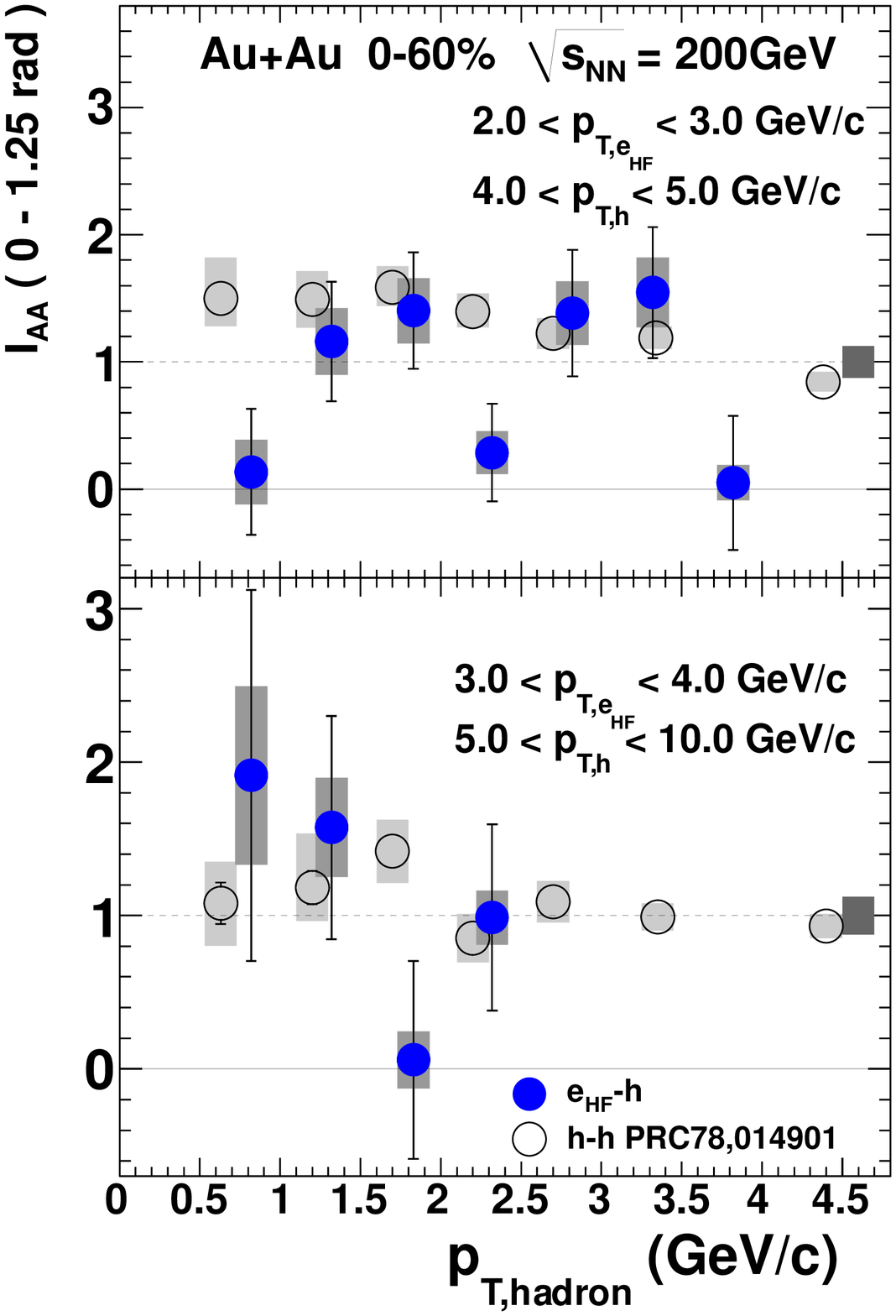}
\caption{\label{iaa_near} (color online) 
Near-side (0$<\Delta\phi<$1.25 rad) $I_{\rm AA}$ for 
2.0$<p_{T,e}<$3.0~GeV/$c$ (top panel) and 3.0$<p_{T,e}<$4.0~GeV/$c$ 
(bottom panel) as a function of the associated hadron $p_T$ for 
$e_{\rm HF}$ (solid points) and hadron (open points) triggers (from 
Ref.~\cite{ppg083}).  The gray band around unity shows the overall 
normalization uncertainty (12.4\%), which moves all points together.  
Points are slightly shifted horizontally for clarity. Figure is from Ref.~\cite{ppg112}.}
\end{minipage}%
 \end{figure*}

The near side yields conditional yields are shown in Figure~\ref{nearyield_auau} 
and are consistent with those observed in p+p
collisions.  
This is quantified by the ratio of the conditional yield in Au+Au collisions
divided by the conditional yield in p+p collisions, $I_{AA}$, shown in
Figure~\ref{iaa_near}.
This is a good cross check because those yields are expected
to be dominated by the decay of the heavy meson ($D$ or $B$)  which occurs far
from the matter and thus should be unaffected by its presence. It is perfectly possible
for the matter to enhance the near side correlations through some process, it is
difficult to imagine a scenario where the yields are suppressed.

\begin{figure*}[h]
\begin{minipage}{0.6\linewidth}
\includegraphics[width=1.00\linewidth]{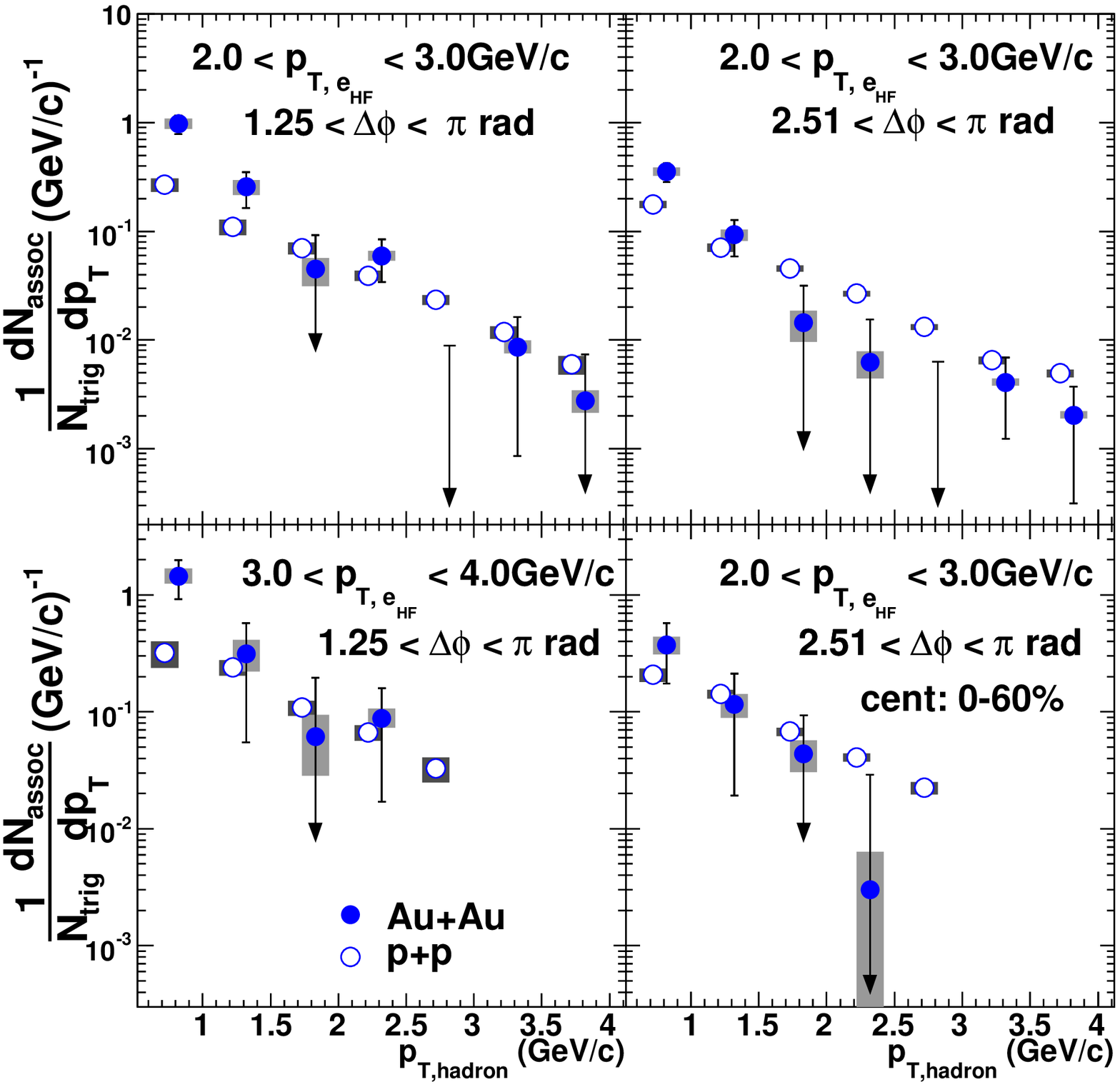}
\end{minipage}
\begin{minipage}{0.35\linewidth}
\caption{\label{yields_away_2panel}
Away-side conditional yields for wide (left) and narrow 
(right) away-side $\Delta\phi$ integration ranges for Au+Au (solid 
points) and $p$+$p$ (open points).  Top panels show 
2.0$<p_{T,e}<$3.0~GeV/$c$ and bottom panels shown 
3.0$<p_{T,e}<$4.0~GeV/$c$.  Upper limits are for 90\% confidence 
levels.  The overall normalization uncertainty of 9.4\% in Au+Au and 
7.9\% in $p$+$p$ are not shown.  Points are slightly shifted 
horizontally for clarity. Figure is from Ref.~\cite{ppg112}.}
\end{minipage}
\end{figure*}

\begin{figure*}[h]
\begin{minipage}{0.6\linewidth}
\includegraphics[width=1.0\linewidth]{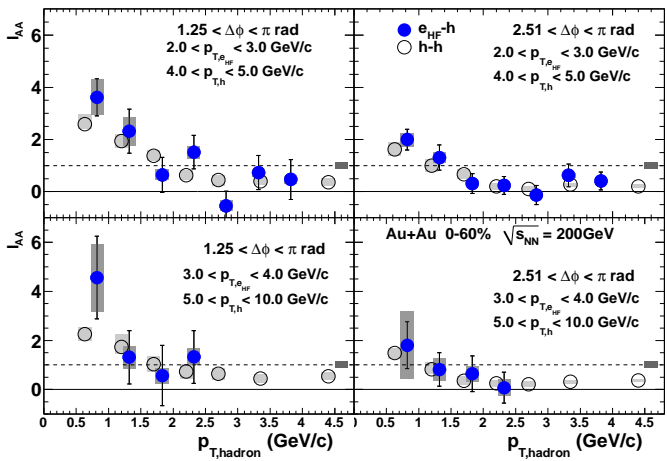}
\end{minipage}
\begin{minipage}{0.35\linewidth}
\caption{\label{away_2panel} 
$I_{\rm AA}$ as determined from the away-side yields in 
Fig.~\ref{yields_away_2panel}.  Two $\Delta\phi$ ranges are shown: 
1.25$<\Delta\phi<\pi$~rad (left panels) and 2.51$<\Delta\phi<\pi$~rad 
(right panels).  The gray band around unity shows the overall 
normalization uncertainty of 12.4\%, which moves all points together.  
For comparison hadron-hadron $I_{\rm AA}$ values from 
Ref.~\cite{ppg083} are also shown for trigger $p_T$ selections where 
the parent heavy meson has similar $p_T$ to the trigger light hadron 
(see Table~\ref{tabpt}).  Points are slightly shifted horizontally 
for clarity.  The solid horizontal line is at 0 and the dashed 
horizontal line is at 1. Figure is from Ref.~\cite{ppg112}.}
\end{minipage}
\end{figure*}

Figure~\ref{yields_away_2panel} shows the away side conditional yields for both Au+Au and
p+p collisions as a function of the associated hadron $p_T$ and for two different 
trigger $p_T$ selections and two ranges for the away side azimuthal angle integration.
The correspondig $I_{AA}$ values are shown in Figure~\ref{away_2panel}.  In 
general at low $p_{T,hadron}$ the yields are enhanced in Au+Au collisions compared
to p+p and at higher $p_{T,hadron}$ the yields are suppressed.  Qualitatively,
this is consistent to trends found in hadron-hadron correlations (see e.g. Ref.~\cite{ppg083}).
In order to make a more quantiative comparison we take into account that the 
electron $p_T$ is not the relevant $p_T$ with which to compare to the hadron-hadron
$I_{AA}$ values.  We use PYTHIA~\cite{pythia} simulations of the parent $B$ and $D$ (weighted
according to Fixed Order Next to Leading Log (FONLL)~\cite{fonll} 
calculations of the mixture of charm
and bottom) $p_T$s 
which give rise to the electrons in our trigger $p_T$ selections
(see Table~\ref{tabpt}) and compare to
hadron-hadron $I_{AA}$ at comparable $p_T$ values.  
This comparison is shown in Figure~\ref{away_2panel}; good agreement is
seen between the electron-hadron and hadron-hadron $I_{AA}$ values.
It might also be reasonable
to compare the hadron-hadron and electron-hadron $I_{AA}$ at 
similar parton $p_T$, accounting for the different fragmentation of
heavy and light quarks.  However, given the current uncertainties, the
comparison at similar meson $p_T$ seems reasonable and gives good agreement.


\begin{table*}[ht]
\begin{tabular}{ccccc}
$ p_{T,e}$ (GeV/$c$) & $\langle p_T \rangle_{D}$ (GeV/$c$)&
$\langle p_T \rangle_{B}$ (GeV/$c$) & $\frac{b\to e}{(c\to e + b\to e)}$ & $\langle p_T \rangle_{meson}$ (GeV/$c$)\\
\hline
1.5-2.0 & 3.4 & 4.4 & 0.15 & 3.6\\
2.0-3.0 & 4.1 & 4.7 & 0.26 & 4.3\\
3.0-4.0 & 5.6 & 5.6 & 0.42 & 5.6\\
\end{tabular} 
\caption{Mean transverse momentum of the parent $D$ and $B$ mesons 
contributing to the heavy-flavor electron $p_T$ bins used here.  
They are combined according to the fraction of heavy-flavor electrons 
from $b$ quarks, $\frac{b \to e} {(c\to e + b\to e)}$ according to 
the FONLL calculations~\cite{fonll} (as shown in Ref.~\cite{ppg094}) 
to determine the mean heavy meson transverse momentum.}
\label{tabpt}
\end{table*}

For many years, a strongly modified away side shape has been seen in 
hadron-hadron correlations~\cite{ppg032}.  This has been attributed to
various sources, among the most prevelant being the remains of a Mach
cone~\cite{jorge}.  With that motivation in mind we analyzed the away 
side shape for electron-hadron correlations.  Heavy quarks, especially
bottom, at moderate $p_T$ values travel at a velocity slower than the speed
of light.  Since the angle of the Mach cone opening depends on the velocity
of the parton relative to the speed of sound, bottom quarks offer the opportunity
to confirm a Mach cone by looking for a variation in the opening angle.

\begin{figure}[ht]
\begin{minipage}{0.6\linewidth}
\includegraphics[width=1.0\linewidth]{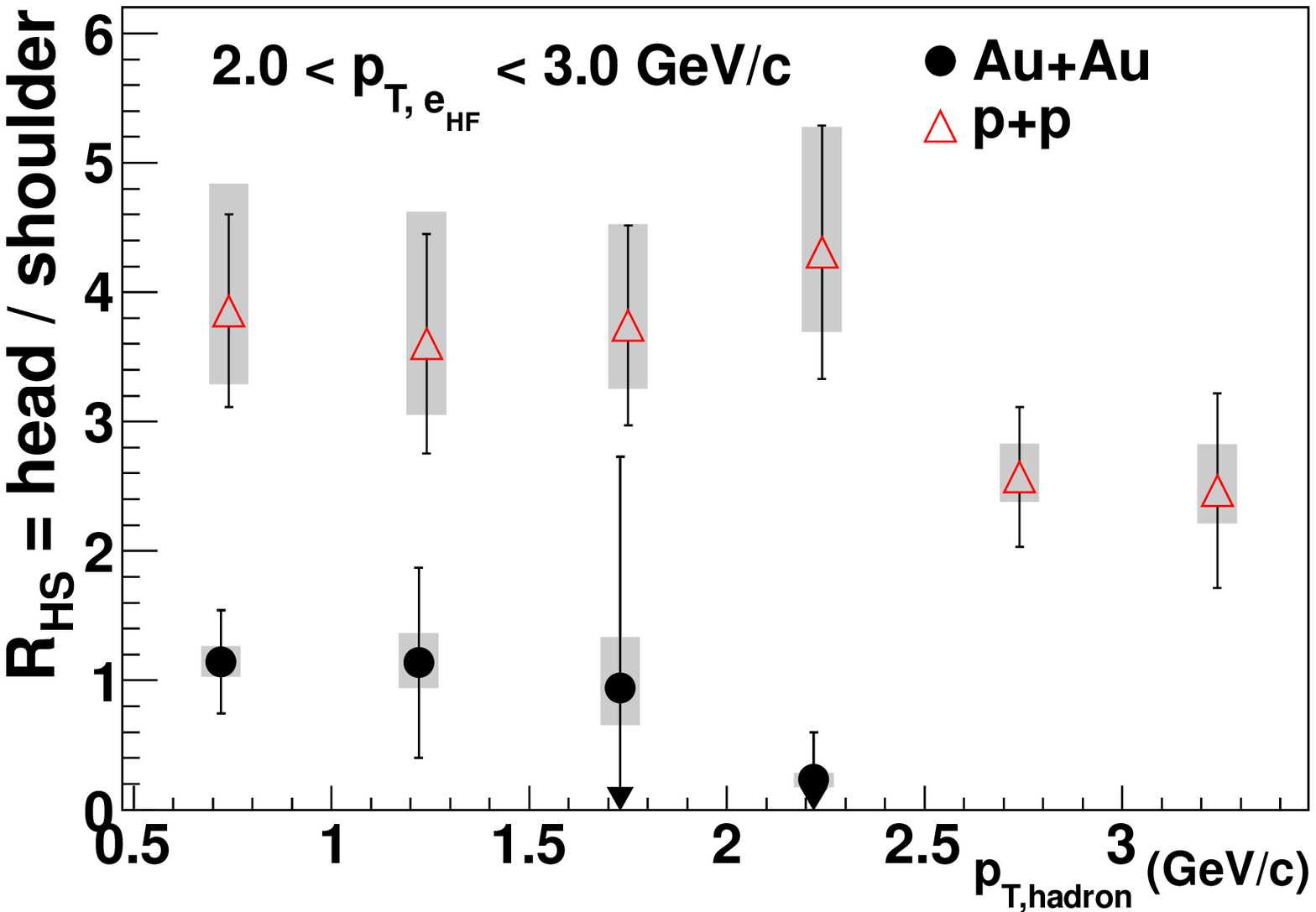}
\end{minipage}
\begin{minipage}{0.35\linewidth}
\caption{Ratio of the yield in the head region per radian to that in 
the shoulder region per radian for Au+Au (black) and $p$+$p$ (red).
Figure is from Ref.~\cite{ppg112}.}
\label{HS_ratio}
\end{minipage}
\end{figure}

We measured the head-to-shoulder ratio, $R_{HS}$, which is the yield per
radian measured into the head region, 2.51$<\Delta\phi<\pi$ rad,
divided by the same quantity in the shoulder region, 1.25$<\Delta\phi<$2.51rad.
The results for p+p and Au+Au collisions are shown in Figure~\ref{HS_ratio}.
There is a significant increase in this ratio in Au+Au collisions compared
to p+p collisions.  No dependence on the associated particle $p_T$ is observed.

Since the publication of these data~\cite{ppg112}, there is evidence that the modified away
side structure is due to fluctuations in the initial state of the Au+Au 
collisions which are then transported to the final state hydrodynamically~\cite{alver,ppg132,alice}.  
The modification of the away side correlations in electron-hadron correlations
seen here is consistent with that understanding.  In the trigger $p_T$ 
selection used in Figure~\ref{HS_ratio}, previous measurements already
show a large $v_2$ value~\cite{ppg066}.  Thus, it would be expected
that higher order Fourier coefficients (e.g. $v_3$), by which these fluctuations have
been measured, would also be seen for these electrons (such a measurement
has not been done, but would be very interesting). 

\section{Conclusions}
The mechanisms by which hard probes interact with the QGP are still not
understood.  Measurements involving heavy quarks have posed a particular
problem to energy loss models based on radiative energy loss.
However, there are many uncertainties in the theoretical calculations
and the need for more and better measurements both at RHIC and the LHC.
At PHENIX, the Silicon Vertex Detector was installed for the 2011 RHIC
running period which will allow the separation of electrons from charm and bottom.
This will be a major step forward to understand how the parton mass alters the
effects of parton-matter interactions.

\section*{References}
\bibliographystyle{iopart-num}
\bibliography{sickles_wwnd11_proceedings}

\end{document}